\documentclass[reprint,superscriptaddress,aps,prb]{revtex4-1}
\usepackage{amsmath}
\usepackage{graphicx}
\usepackage{units}
\begin{document}
\title{Revealing weak spin-orbit coupling effects on charge carriers in a $\pi$-conjugated polymer}
\author{H.~Malissa}
\email{hans.malissa@utah.edu}
\affiliation{Department of Physics and Astronomy, University of Utah, Salt Lake City, UT 84112, USA}
\author{R.~Miller}
\affiliation{Department of Physics and Astronomy, University of Utah, Salt Lake City, UT 84112, USA}
\author{D.~L.~Baird}
\affiliation{Department of Physics and Astronomy, University of Utah, Salt Lake City, UT 84112, USA}
\author{S.~Jamali}
\affiliation{Department of Physics and Astronomy, University of Utah, Salt Lake City, UT 84112, USA}
\author{G.~Joshi}
\affiliation{Department of Physics and Astronomy, University of Utah, Salt Lake City, UT 84112, USA}
\author{M.~Bursch}
\affiliation{Mulliken Center for Theoretical Chemistry, Institut f{\"u}r Physikalische und Theoretische Chemie, Universit{\"a}t Bonn, 53113 Bonn, Germany}
\author{S.~Grimme}
\affiliation{Mulliken Center for Theoretical Chemistry, Institut f{\"u}r Physikalische und Theoretische Chemie, Universit{\"a}t Bonn, 53113 Bonn, Germany}
\author{J.~van~Tol}
\affiliation{National High Magnetic field Laboratory, Florida State University, Tallahassee, FL 32310, USA}
\author{J.~M.~Lupton}
\affiliation{Department of Physics and Astronomy, University of Utah, Salt Lake City, UT 84112, USA}
\affiliation{Institut f{\"u}r Experimentelle und Angewandte Physik, Universit{\"a}t Regensburg, Universit{\"a}tsstrasse 31, 93053 Regensburg, Germany}
\author{C.~Boehme}
\affiliation{Department of Physics and Astronomy, University of Utah, Salt Lake City, UT 84112, USA}
\date{\today}
\begin{abstract}
We measure electrically detected magnetic resonance (EDMR) on organic light-emitting diodes (OLEDs) made of the polymer poly[2-methoxy-5-(2-ethylhexyloxy)-1,4-phenylenevinylene] (MEH-PPV) at room temperature and high magnetic fields, where spectral broadening of the resonance due to spin-orbit coupling (SOC) exceeds that due to the local hyperfine fields. Density-functional-theory calculations on an open-shell model of the material reveal $g$-tensors of charge-carrier spins in the lowest unoccupied (electron) and highest occupied (hole) molecular orbitals. These tensors are used for simulations of magnetic resonance line-shapes. Besides providing the first quantification and direct observation of SOC effects on charge-carrier states in these weakly SO-coupled hydrocarbons, this procedure demonstrates that spin-related phenomena in these materials are fundamentally monomolecular in nature. 
\end{abstract}
\maketitle
Charge-carrier states of hydrocarbon-based materials such as $\pi$-conjugated polymer films are known to exhibit very weak spin-orbit coupling (SOC) compared to many other compounds \cite{Liang2016, Yu2012, Uoyama2012, Yu2011, Barford2010, Rybicki2009, McClure1952}. For the description of some of these materials' physical behavior, including their magnetoresistance, luminescence, and permeability \cite{Liang2016, Yago2011, Beljonne2001, Wakasa1999}, it is a reasonable assumption to consider SOC to be negligible \cite{Barford2010}. In contrast to this, however, there are other material properties such as high-field magneto-optoelectronic characteristics \cite{Schellekens2011, Gladkikh2006}, spin lifetimes \cite{Gladkikh2006, Nuccio2013}, and spin diffusion lengths \cite{Liang2016, Yu2011}, the inverse spin-Hall effect \cite{Kavand2017, Sun2016, Yu2015, Ando2013} or the general spin statistics of organic light-emitting diodes (OLEDs), which determine overall OLED efficiency, where this assumption leads to incorrect predictions. Furthermore, the study of the influence of SOC on paramagnetic states---i.\ e.\ on polarons---such as $g$-factor shifts and anisotropies opens up a route to scrutinizing theoretical predictions of the nature of these states. Quantum chemistry is used to calculate energy levels by computing the carrier wave functions. The quality of these calculations can be tested by examining the $g$-tensors computed by density-functional-theory (DFT) methods. There have been recent material studies which specifically aim at the investigation and control of SOC in organic semiconductors \cite{Schott2017}. However, these studies are based on doping with heavy elements, which induce strong SOC, and are only peripherally of relevance to organic OLEDs since they are carried out in solution.

In the following we investigate the effects of SOC on the Land\'e $g$-factors \cite{Segal1965} of charge carriers in low-lying electronic cationic and anionic states [as approximated by the lowest unoccupied molecular orbital (LUMO) and the highest occupied molecular orbital (HOMO), respectively] containing mobile electrons or holes, in the $\pi$-conjugated polymer poly[2-methoxy-5-(2-ethylhexyloxy)-1,4-phenylenevinylene] (MEH-PPV). We consider thin films of the material in OLED devices under room-temperature bipolar charge-carrier injection conditions. The experimental approach taken here is based on the shifts and broadening of magnetic resonance spectra, which arise due to the influence of SOC on the $g$-factor. Since MEH-PPV consists of relatively light elements (C, H, and O), SOC is weak and $g$-factors are therefore close to the free-electron value of 2.002319 \cite{Joshi2016, Baker2012, McCamey2008}. However, because even weak SOC leads to a minuscule $g$-factor shift and gives rise to spectroscopic fingerprints such as an anisotropic $g$-tensor, it becomes detectable in magnetic resonance spectroscopy at large $B_0$ fields, i.\ e.\ at large Zeeman splitting of the spin levels. To detect SOC effects in magnetic resonance, the spectral broadening induced by SOC must exceed that arising from local hyperfine fields due to hydrogen nuclei \cite{Joshi2016, Waters2015, Malissa2014}. 

\begin{figure}
\includegraphics[width=\columnwidth]{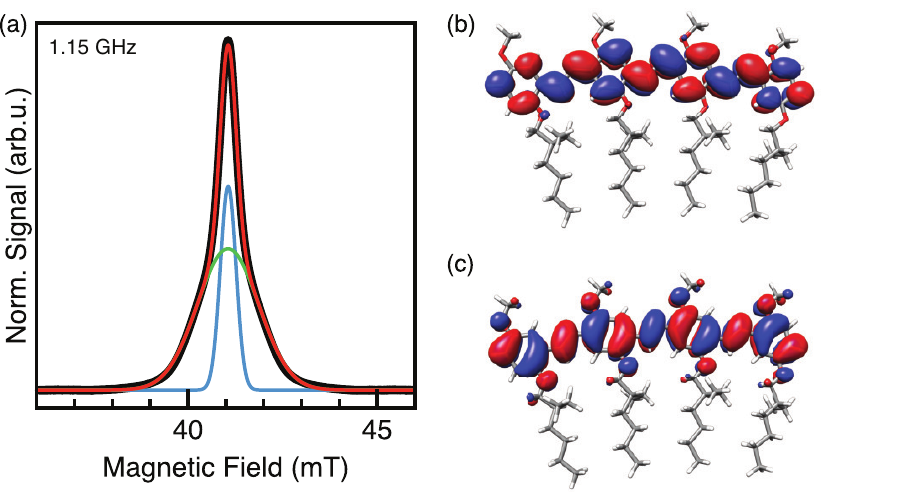}
\caption{\label{FigureIntro} (a) Room-temperature EDMR spectrum of MEH-PPV with a double-Gaussian fit (red). (b, c) Plots of the single-carrier probability density in the LUMO (b) and HOMO (c) of a model of MEH-PPV.}
\end{figure}

We therefore use high-magnetic-field \cite{vanTol2005}
electrically detected magnetic resonance spectroscopy (EDMR), which enables the measurement of charge-carrier (polaron) pair magnetic resonance spectra of electrons and holes by recording spin-dependent electrical recombination currents in OLEDs \cite{Kavand2016,Joshi2016,Malissa2014,Baker2012,McCamey2008}. 
We note that extensive previous work has demonstrated that the resonant species are indeed weakly spin-spin coupled $S=\nicefrac{1}{2}$ carriers \cite{Waters2015,vanSchooten2015,McCamey2008}. This information derives from time-resolved measurements of the Rabi precession frequency \cite{Waters2015,McCamey2008}. The minuscule zero-field splitting of the carrier pair corresponds to fields of approximately \unit[100]{nT} \cite{vanSchooten2015} and is therefore entirely irrelevant for the high-field experiments discussed here. We have discussed technical details of low-field EDMR experiments on such devices in detail previously \cite{Joshi2016, Baker2012, Lee2012}, but all these reports pertained to EDMR experiments below fields of \unit[1]{T}. Under these low-field conditions, $g$-factor distributions in MEH-PPV due to weak SOC are spectroscopically detectable only qualitatively above fields of \unit[300]{mT} in the form of a magnetic-field dependent spectral broadening \cite{Joshi2016}. However, on an absolute scale, at these low fields, SOC effects are negligible compared to the contributions to the resonance spectrum from strong hyperfine field distributions of the omnipresent hydrogen nuclei. Fig.~\ref{FigureIntro}(a) shows an EDMR spectrum of an MEH-PPV OLED measured at a microwave (MW) frequency of \unit[1.15]{GHz} using $B_0$-modulation and lock-in detection \cite{Joshi2016}. Note that the displayed data represent the integrated signal. At this low MW frequency, the line widths of electron and hole resonances are governed solely by the hyperfine coupling with surrounding hydrogen nuclei, which give rise to Gaussian disorder broadening. The observed spectra can therefore be modeled by a superposition of two Gaussian lines with identical line centers, i.\ e.\ with effective $g$-factors of free electrons, but with two different line widths. From the fit shown in Fig.~\ref{FigureIntro}(a), spectral widths of \unit[0.208]{mT} and \unit[0.811]{mT} are obtained, which provide a quantification of the distribution in hyperfine field strengths. 

In contrast to the data of Fig.~\ref{FigureIntro}(a), the high-field EDMR experiments reported here allow us to resolve shifts in the $g$-factor of the charge-carrier spins probed arising due to the small but finite SOC \cite{Fehr2011}. These shifts become relevant on magnetic field scales which significantly exceed the magnetic-field range of hyperfine broadening. In particular,  high-field EDMR measurements reveal anisotropic $g$-tensors which, together with the random spatial orientation of orbitals within the ensembles, give rise to asymmetries in the resonance lines. 

In order to interpret the experimental results, we conducted DFT quantum-chemical calculations on a spin-polarized open shell ($S=\nicefrac{1}{2}$) model of the material by using molecular structures consisting of four monomer units of MEH-PPV with either an electron added (MEH-PPV$^-$) or removed (MEH-PPV$^+$). The structures were optimized in the gas phase at the TPSSh-D3(BJ)/def2-TZVP \cite{Grimme2016, Grimme2011, Grimme2010, Grimme2007, Weigend2005, Staroverov2003} level of theory using the TURBOMOLE program package \cite{TURBOMOLE, Furche2014}. The computed molecular orbitals (MOs) of the electron (LUMO) and hole (HOMO) based on the optimized structures are shown in Fig.~\ref{FigureIntro}(b,c) \cite{Chen2014}. DFT calculations further yield electronic $g$-tensors for both model systems at the TPSSh/IGLO-III/TPSSh-D3(BJ)/def2-TZVP \cite{Kutzelnigg1990} level of theory provided by the ORCA program package \cite{Neese2012}. 

\begin{figure}
\includegraphics[width=\columnwidth]{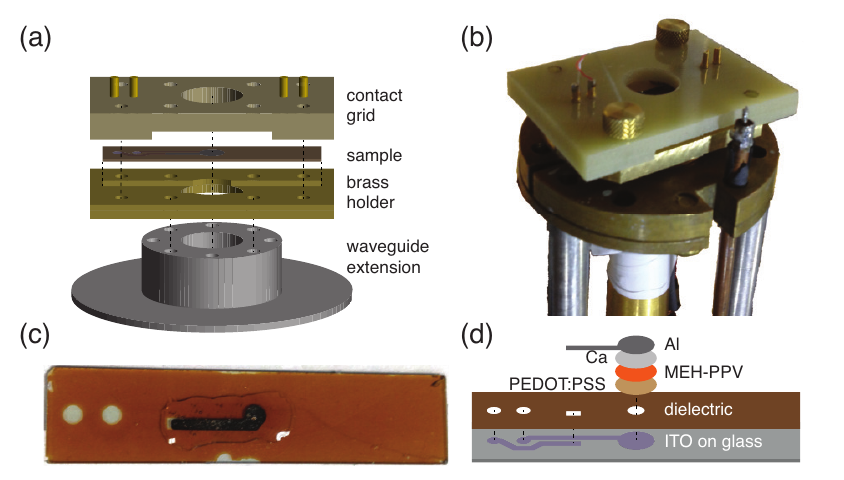}
\caption{\label{FigureSample} (a) Sketch of the high-magnetic-field EDMR probe head. (b) Photograph of the probe head when mounted to the waveguide. (c) The OLED sample for high-field EDMR. (d) Sketch of the sample design (the encapsulation is omitted).}
\end{figure}

EDMR experiments at very high MW frequencies \cite{Ling2014, Fehr2011} were carried out using a 120, 240, and \unit[336]{GHz} multi-frequency quasi-optical heterodyne EPR spectrometer at the National High Magnetic Field Laboratory of Florida State University \cite{vanTol2005}. In this setup, the sample is irradiated with mm waves while mounted at the end of a corrugated waveguide. We developed a sample holder which allows for electrical connection to the OLED while meeting the geometrical requirements of the spectrometer's waveguide structures. The sample and sample-holder designs are shown in Fig.~\ref{FigureSample}. Fig.~\ref{FigureSample}(a) sketches the sample holder which encompasses a waveguide extension and mounting flange, a brass plate with a rectangular pocket to accommodate the sample, and a bore hole in order to allow for MW irradiation. The OLED sample was fabricated on a rectangular glass template of dimensions $\unit[44\times 10]{mm^2}$, with the active area at the center, and aligned with the waveguide and a fiberglass spacer with a spring-loaded gold contact plate to keep it in place and provide the electrical contact. Fig.~\ref{FigureSample}(b) shows the assembled sample holder mounted to the corrugated waveguide of the spectrometer. The sample [cf.\ Fig.~\ref{FigureSample}(c)] has lithographically defined indium-tin oxide thin-film wiring which defines the OLED back contact and the contact pads. A layer of hard-baked photoresist with apertures serves as an electrical insulator and defines the circular OLED pixel with a diameter of \unit[2]{mm}.
Fig.~\ref{FigureSample}(d) shows the lateral and vertical OLED structure: a layer of PEDOT:PSS serves as a hole injector, and a layer of MEH-PPV constitutes the active device. Layers of Ca and Al form electron injection layers and electrical interconnects to the contact pads, respectively. The device structure is analogous to that described in Ref.~\onlinecite{McCamey2008}, with a geometrical arrangement which is more suitable for this particular application where the sample is irradiated with mm waves. 

\begin{figure}
\includegraphics[width=\columnwidth]{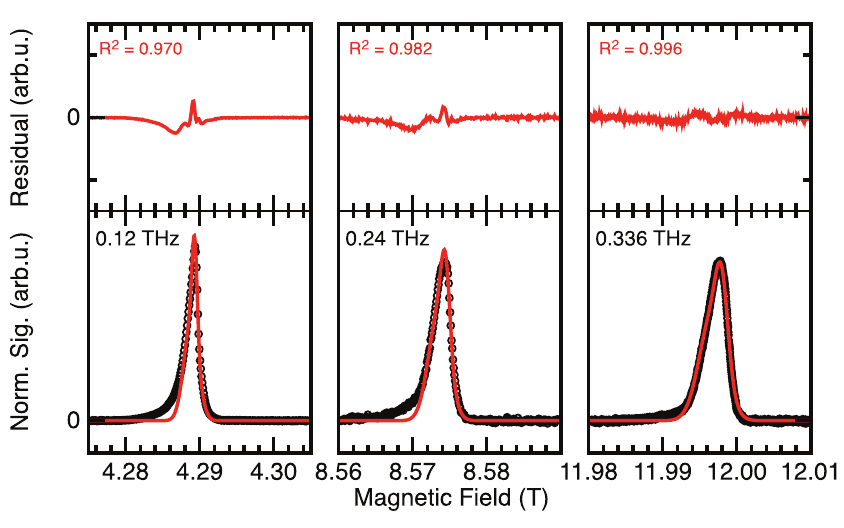}
\caption{\label{FigureGauss} Lower panels: EDMR spectra at MW frequencies of 120, 240, and \unit[336]{GHz}, along with a global fit to a double-Gaussian model with frequency-dependent broadening. Upper panels: residuals, along with coefficients of determination.}
\end{figure}

\begin{figure}
\includegraphics[width=\columnwidth]{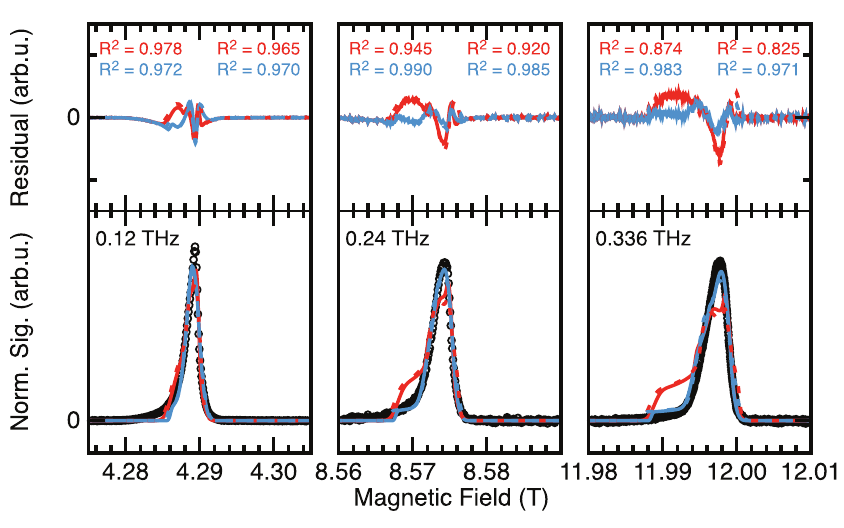}
\caption{\label{FigureSim} Measured and simulated spectra. Red lines represent simulations where the narrow line width is assigned to the electron and the broad line width to the hole. Blue lines show opposite assignment. Solid lines correspond to a geometry where the substituents are located on alternating sides of each monomer unit, whereas dashed lines show the all-\emph{trans} conformation.}
\end{figure}

The results of amplitude-modulated high-field EDMR experiments are summarized in Figs.~\ref{FigureGauss} and \ref{FigureSim}. For these experiments, a forward bias of \unit[3.5]{V} was applied to the sample to give a $\unit[20]{\mu A}$ current. The EDMR spectra show resonance peaks with substantially broadened line shapes and strong deviations from the symmetric double-Gaussian line which is characteristic of the EDMR experiments at lower MW frequencies \cite{Joshi2016}. In Fig.~\ref{FigureGauss}, the measured spectra are shown along with a least-squares fit to a global model which assumes a line width for each of the two charge carriers $\sigma=\sqrt{\Delta B_{\text{hyp}}^2+\alpha^2B_0^2}$, a geometric sum of a constant term ($B_0$-independent hyperfine fields $\Delta B_{\text{hyp}}$) and a broadening term scaling linearly with $B_0$ \footnote{Note that the line width is defined as the Gaussian root-mean-square width $\sigma$ rather than the full width at half maximum (FWHM) throughout this work as well as in Ref.~\onlinecite{Joshi2016} unless stated otherwise. $\sigma$ and the FWHM are related according to $\mbox{FWHM}=2\sqrt{2\ln 2}\sigma$.}. This so-called $g$-strain broadening assumes a Gaussian distribution of $g$-factors due to molecular structural disorder which modifies the strength of SOC across the sample. This broadening is isotropic and is described by a $B_0$-dependent Gaussian line width contribution $\alpha B_0$ (with $\alpha$ a dimensionless scaling parameter), as described in Refs.~\onlinecite{Joshi2016}. Differences in the $g$-factors of electrons and holes correspond to an offset between the centers of the two Gaussian lines which scales linearly with $B_0$. While such an asymmetry of resonance line centers is not observed at frequencies $\unit[<20]{GHz}$ where the line width dominated by hyperfine fields \cite{Joshi2016} exceeds by far any offset induced by differences in $g$-factor between the resonance lines, it may play a role in the high-field regime. 

Using this global model to fit the measured spectra together with the \unit[1.15]{GHz} spectrum [cf.\ Fig.~\ref{FigureIntro}(a)] we obtain line shapes which describe all three high-field spectra to a certain extent. The free global parameters in this fit are the $g$-factors of both charge carriers, the two constant line widths, and the two field-dependent broadening terms $\alpha$ \cite{Joshi2016}. The upper panels in Fig.~\ref{FigureGauss} show the residuals of this fit, along with coefficients of determination. From this least-squares fit we obtain the parameters for the two charge carriers $g_1=2.002680\pm 0.000005$, $\Delta B_{\text{hyp},1}=\unit[0.2080\pm 0.0008]{mT}$, $\alpha_1=\left(6.735\pm 0.044\right)\times 10^{-5}$ for the narrow resonance and $g_2=2.002906\pm 0.000005$, $\Delta B_{\text{hyp},2}=\unit[0.8111\pm 0.0027]{mT}$, $\alpha_2=\left(1.258\pm 0.008\right)\times 10^{-4}$ for the wider line. The respective hyperfine field line width contributions $\Delta B_{\text{hyp}}$ are determined by the low-frequency spectrum in Fig.~\ref{FigureIntro}(a), and while they are similar to the ones given in Ref.~\onlinecite{Joshi2016}, they lie outside of the stated 95\% parameter confidence interval. More significantly, the values for the SOC induced broadening parameter $\alpha$ are much smaller than those previously extrapolated in Ref.~\onlinecite{Joshi2016} ($\alpha_1$ and $\alpha_2$ are reduced by approximately 60\% and 75\%, respectively). Thus, from the fit results in Fig.~\ref{FigureGauss}, we conclude that the simple global model of a double Gaussian line shape yields only limited agreement with the data. Furthermore, the data substantially contradict the frequency dependence of the line widths determined in Ref.~\onlinecite{Joshi2016}, where it was assumed that SOC leads to isotropic broadening with a line-width term that scales linearly with $B_0$. We conclude that, while for the magnetic field domain below \unit[700]{mT} studied in Ref.~\onlinecite{Joshi2016} this simple model is sufficient, i.\ e.\ it is well suited for the accurate determination of hyperfine-field distributions of the two charge-carrier types, the assumption of isotropic $g$-tensors (i.\ e.\ the applicability of scalar $g$-factors) is insufficient at higher magnetic fields where SOC contributes to the spectrum. 

An appropriate description of the spectra at all frequencies can only be made by performing DFT calculations to assess the $g$-factor anisotropy. We use the calculated $g$-tensors together with the constant isotropic line widths and peak-area ratios obtained in the low-field regime [cf.\ Fig.~\ref{FigureIntro}(a)] to simulate the spectrum using the EasySpin toolbox \cite{Stoll2006}. The results of this procedure, along with the spectra, are shown in Fig.~\ref{FigureSim}. The red lines correspond to cases where the narrow hyperfine distribution [\unit[0.5]{mT} half width, cf.\ Fig.~\ref{FigureIntro}(a)] is assigned to the electron and the broad distribution (\unit[1.9]{mT} half width) to the hole. The blue curves describe the opposite case. In addition to this distinction, the $g$-tensors are calculated separately for different molecular geometries, i.\ e.\ the orientation of the side chain substituents of MEH-PPV is considered. The solid curves correspond to a polymer geometry where the side groups are located on alternating sides for each monomer unit, whereas the dashed lines are computed for an arrangement with all chains lying in parallel, the all-\emph{trans} configuration [cf.\ Fig.~\ref{FigureIntro}(b,c)]. We emphasize that these simulated spectra are fitted to the experimental data merely by adjustment of a linear vertical scaling factor to compensate for the arbitrary signal amplitude and a horizontal offset to account for absolute measurement error in $B_0$ at high magnetic field. The anisotropic line shape is purely a result of the computed $g$-tensor anisotropy and the experimentally determined hyperfine field line widths. 
The upper panels in Fig.~\ref{FigureSim} show the fit residuals of the different models, along with the coefficients of determination of each curve \footnote{Note that the $g$-tensors used for the line shape simulation result from the DFT calculation and not from the least-squares fit. Confidence limits can therefore not be estimated from the line shape simulation. There is only a limited amount of literature that pertains to the comparison of similar calculations to measurments (c.\ f., for example, Ref.~\onlinecite{Kaupp2002}).}. The blue curves are in good agreement with the measurements, and the overall line width and shape are reproduced for all MW frequencies. In contrast, the red curves deviate substantially for higher MW frequencies. The simulated line is much broader than the experimental result and exhibits a bifurcation which is not observed in experiment. The simulation therefore indicates that the blue line resembles the physical reality more closely than the red line: the narrow hyperfine distribution is therefore experienced by the hole spin and the broad distribution by the electron. In other words, the hole spin is less strongly hyperfine coupled than the electron spin; the hole wave function is therefore more delocalized, while exhibiting a larger $g$-factor shift due to SOC.

The differences between solid and dashed curves in Fig.~\ref{FigureSim} are not very pronounced, implying that \emph{cis-trans} isomerization has little influence on SOC. The coefficients of determination for the dashed spectra [all side chains parallel, cf.\ Fig.~\ref{FigureIntro}(b,c)], stated in the figure, are marginally better than for the solid curves. These two situations represent extreme cases of molecular ordering, and a disordered arrangement of the monomers is much more realistic. No further frequency-dependent broadening mechanisms, such as those due to material inhomogeneity and disorder as discussed in the  model in Ref.~\onlinecite{Joshi2016}, are taken into account here. Temporal fluctuations of the $g$-tensors due to molecular dynamics would lead to an even stronger frequency-dependent broadening, and the fact that the experimentally observed line widths are in such good agreement with simulation indicates that this effect, though conceivable, must be small. 

From the comparison of the measured and calculated EDMR spectra we conclude that the computed polaron molecular orbitals and $g$-tensors provide an accurate description of the physical reality and, even though the simulation is only based on non-interacting segments of four monomer units, this approximation is basically applicable to device operating conditions of the thin films. High-field EDMR therefore offers a unique probe of the polaron wave functions in organic semiconductors, demonstrating that these are truly monomolecular in nature: there is no significant delocalization of the wave function and intermolecular packing does not appear to influence the $g$-tensor. We also note that we can exclude the influence of any further resonant paramagnetic species. Although it is possible that charge trapping states can influence steady-state magnetic-field effects such as magnetoresistance \cite{Cox2014a,Cox2014}, such states would appear in the resonance spectrum. However, in this case, we would not be able to find such a high level of agreement between the calculated $g$-tensors and the measured spectra, since trap states were not explicitly considered in the calculation. Our observation that electrons in MEH-PPV experience stronger random hyperfine fields compared to holes confirms a previously expressed hypothesis based on EDMR spectroscopy of electron-acceptor interface processes in bulk heterojunctions \cite{Matsui2008, Ceuster2001}. In addition, our results suggest that structural disorder has only limited influence on the intermolecular distribution of the magnitude of SOC, a conclusion in agreement with recent qualitative results on the conjugated polymer polyfluorene, a polymer which exists in two distinct structural phases \cite{Miller2016}. These results clearly imply that magnetoresistive and magneto-optical effects in OLEDs of this material, arising due to the distribution in $g$-factors, will only become relevant at high fields\cite{Lupton2008}. We conclude that high-field EDMR offers a powerful route to quantifying the influence of SOC. Given the sensitivity of the technique to small perturbations of SOC, 
this spectroscopy 
can be used
to derive fundamental correlations between molecular structure and SOC. Such correlations are particularly important in unravelling the interplay between enhanced radiative and non-radiative emission by SOC in phosphorescent OLED emitters \cite{Ratzke2016}. Another exciting possibility would be to attempt to reversibly switch SOC---by electric fields or photo- or electrothermal conformational or electronic perturbations---and monitor this with EDMR. 
This approach is not limited to organic semiconductors but should also be applicable to other systems such as molecular magnets. 
\begin{acknowledgments}
This work was supported by the US Department of Energy, Office of Basic Energy Sciences, Division of Materials Sciences and Engineering under Award \#DE-SC0000909. Part of this work was performed at the National High Magnetic Field Laboratory, which is supported by NSF Cooperative Agreement No. DMR-1157490 and the State of Florida. The theoretical work was supported by the DFG in the framework of the Gottfried-Wilhelm-Leibniz Award to S. G.
\end{acknowledgments}
\bibliography{references}
\end{document}